\documentclass[twocolumn,aps,noshowpacs]{revtex4}
\usepackage{psfig,amssymb,graphics,epsf,psfrag}
\usepackage[english]{babel}
\begin{document}
\input epsf

\title{Distribution of current in non-equilibrium diffusive systems and phase
transitions}

\author{T. Bodineau$^{\dag }$ and B. Derrida$^{\ddag}$ }
\affiliation{$\dag$ Laboratoire de Probabilit{\'e}s et Mod{\`e}les Al{\'e}atoires,
CNRS-UMR 7599, Universit{\'e}s Paris VI $\&$ VII,
4 place Jussieu, Case 188, F-75252 Paris, Cedex 05;\\
$^{\ddag}$ Laboratoire de Physique Statistique, Ecole Normale Sup{\'e}rieure,
24 rue Lhomond, 75231 Paris Cedex 05, France  
}

\begin{abstract}
We consider diffusive lattice gases on a ring and analyze 
the stability of their density profiles conditionally to a current deviation. 
Depending on the current, one observes a phase transition between a regime 
where the density remains constant and another regime where the density 
becomes time dependent.
Numerical data confirm this phase transition.
This time dependent profile persists in the large drift limit and 
allows one to understand on physical grounds  the results
obtained earlier for the totally asymmetric exclusion process on a
ring.

\vspace{+0.2in}
{02.50.-r, 05.40.-a, 05.70 Ln, 82.20-w}
\end{abstract}

\date{\today}

 \maketitle

\section{Introduction}
\label{sec: 1}

Stochastic lattice gas  models have been extensively studied recently
as they are among the simplest examples of non-equilibrium systems.
 A powerful  approach to understand their steady state 
was developed by  Bertini, De Sole, Gabrielli, Jona-Lasinio, Landim, 
as a macroscopic fluctuation theory (MFT) which gives,  for large
diffusive systems,  the probability distribution
of  trajectories in the space of density profiles \cite{bdgjl,bdgjl2}.
The MFT relies on the hydrodynamic large deviation theory
\cite{kov,spohn,KL}  which provides estimates for the probability of observing 
atypical space/time density profiles.
It  gives a framework to calculate a large number of properties
of stochastic lattice gas, such as the large deviation functional of the
density profiles.
 Recent developments of the hydrodynamic large deviation theory 
\cite{bdgjl3,bd} enabled to estimate also the large deviations of the current 
through the system.
What the  MFT 
provide are the equations of the time evolution of the most
likely density profile responsible of a given fluctuation.
What it does not provide, in general,  is the solution of these equations 
which would give quantitative predictions for the distribution of the
fluctuations. So far, for the large deviation functional of the density
profiles in the steady state,  the equations  could only be solved  in
a few cases of non-equilibrium systems  with open boundaries (the symmetric exclusion process \cite{bdgjl2},
the Kipnis, Marchioro, Presutti model \cite{bgl,kmp}).  For the SSEP the results of the MFT  were in full agreement
with   the results obtained \cite{dls,dls2,ed} from the exact knowledge of the weights of the
microscopic configurations in the steady state.

In our previous work \cite{bd}, we developed a theory to calculate the large
deviation function of the current through a long one dimensional
diffusive lattice gas in contact at its two ends with two reservoirs
at unequal densities.
Our approach was based on an assumption, the {\it additivity
principle},
which relates the large deviation function (LDF) of the current of a system
to the LDF's of subsystems, when one breaks a large system into large
subsystems. This assumption is in fact  equivalent to the
hypothesis, within the  hydrodynamic large deviation framework, that to observe, for a very long time period $T$, an average current $q=Q_T/T$, 
the system adopts a  profile with a shape,
fixed in time, but of course depending on $q$ (here $Q_T$ is the total
number of particles transfered, say from the left reservoir to the
system during time $T$). The additivity principle allows one to obtain explicit expressions
\cite{bd} for
all the cumulants of the integrated current $Q_T$. 
The predictions of our theory were tested in a few cases \cite{bd} and the results
were found in complete agreement with what was already known or what could be
derived by alternative approaches \cite{ddr,wr,hrs}.

Recently, it was pointed out \cite{bdgjl3} that even if  our predictions \cite{bd} are valid
for some diffusive lattice gas, it might happen  that, to
produce an average current $q$ over a long period of time, the best profile is time-dependent.
One of the goals of the present work is to show that, for a simple example,
the weakly asymmetric exclusion process on a ring,  this indeed   happens
for some range of parameters. 

Let us consider, as we shall do it  in the rest of this paper, the time
evolution  of  a one
dimensional stochastic lattice gas on a lattice of $N$ sites.  According to the
 hydrodynamic formalism \cite{spohn}, a given lattice gas can be characterized by two functions $D(\rho)$ and $\sigma(\rho)$ of its density $\rho$.
One way to define them \cite{bd} is to consider a one dimensional system of
length $N$ connected to reservoirs at its two ends.
For such a lattice
gas, the variance of the total charge $Q_T$ transfered during a long time
$T$ from one reservoir to the other is given, for large $N$,  by definition of
$\sigma(\rho)$ by
\begin{equation}
{\langle Q_T^2 \rangle \over T} = {\sigma(\rho) \over N} 
\label{sigmadef}
\end{equation}
when both reservoirs are at the same density $\rho$.
On the other hand if the left reservoir is at density $\rho + \Delta
\rho$ and the right reservoir at density $\rho$, the average current is
given,  for small $\Delta \rho$, by
\begin{equation}
{\langle Q_T \rangle \over T} = { D(\rho)  \Delta \rho \over N} 
\label{Ddef}
\end{equation}
which is simply Fick's law and defines the function $D(\rho)$.
%These two coefficients are related by the Einstein relation
% $\sigma (\rho) = D (\rho) \chi (\rho)$, where $\chi (\rho)$ 
%is the static compressibility of the lattice gas \cite{spohn}.
In the symmetric simple exclusion process $\sigma(\rho)=\rho(1-\rho)$ and
$D(\rho)=1/2$ \cite{spohn} whereas in the Kipnis, Marchioro, Presutti
model \cite{bgl,kmp}
$\sigma(\rho)=\rho^2$ and $D(\rho)=1/2$.
The effect of a uniform weak electric field of strength $\nu/(2N)$
acting from left to right on the particles is to modify
(\ref{Ddef}) into
\begin{equation}
{\langle Q_T \rangle \over T} = { D(\rho)  \Delta \rho \over N}  + {\nu  \sigma(\rho)   \over N}
\label{Ddef1}
\end{equation}
This equation follows from the linear response theory \cite{spohn}.

Once $D(\rho)$ and $\sigma(\rho)$ are known for a given diffusive system,
the probability of observing the evolution of a density profile $\rho(x,s)$
and a rescaled current $j(x,s)$
for $0 < s < T$ during a time $T \sim N^2$  is given, according to the 
hydrodynamic large deviation theory \cite{bdgjl3},   by
\begin{equation}
{\rm Pro} \big( j(x,s), \rho(x,s) \big) \sim \exp \left[ - 
{{\mathcal{I}}^\nu_{[0,T]}(j,\rho) \over N}\right]
\label{mft}
\end{equation}
where  $\mathcal{I}_{[0,T]}^\nu$ is defined by
\begin{eqnarray}
\lefteqn{
\mathcal{I}_{[0,T]}^\nu(j,\rho) 
= \int_0^T ds \int_0^1 dx  
}\label{Idef}
                    \\
& {\displaystyle {[j(x,s) + D(\rho(x,s)) \rho'(x,s) - \nu \sigma(\rho(x,s))]^2 \over
2 \sigma(\rho(x,s))}}  \nonumber
\end{eqnarray}
with $\rho' = \partial  \rho/ \partial x$
and where the rescaled current $j(x,s)$ 
 is related to the density profile
$\rho(x,s)$ by the
conservation law
\begin{equation}
 {d \rho(x,s) \over ds} = - {d j(x,s) \over dx} 
\label{conservation}
\end{equation}
 A formalism equivalent to this hydrodynamic large deviation theory was developed independently \cite{pjsb,jsp} in the context of the full counting statistics of the
transport of free fermions through disordered wires.
 A simple derivation of (\ref{Idef}) and (\ref{conservation}) is given in the appendix.
\medskip

The large deviation function $G(j_0)$ of the current is then    defined as
\begin{eqnarray}
\label{proj0}  
{\rm Pro}\left( {Q_T \over T} = {j_0 \over N}  \right) 
\sim   \exp \left( {T \over N} G(j_0) \right)\\
\text{for large $T$ and $N$} \nonumber  
\end{eqnarray}
(In (\ref{proj0}), one has first to take the limit $T \to \infty$ and
then make $N$ large; in practice (\ref{proj0}) should hold when $T \gg
N^2$ as  $N^2$ is the characteristic time  of a diffusive system of size $N$).

Now according to (\ref{mft}), the large deviation function $G(j_0)$ is given by
\begin{equation} G(j_0) = \lim_{T \to \infty}
 \left[ - {1 \over T} 
 \min_{\rho(x,s)} 
\mathcal{I}_{[0,T]}^\nu(j,\rho) \right]
\label{Gj0}
\end{equation}
where the current $j(x,s)$ satisfies for large $T$ and all $x$ 
the constraint
\begin{equation}
 \lim_{T \to \infty} {1 \over T}\int_0^T  j(x,s) ds = j_0 
\label{constraint}
\end{equation}
with the profile $\rho(x,s)$ and the current $j(x,s)$  connected by
(\ref{conservation}).

In the following, we will often consider, instead of (\ref{proj0}), the
generating function of the current:
\begin{equation}
 \left\langle e^{\lambda Q_T} \right\rangle \sim e^{T \mu(\lambda)}  \ \ \ {\rm for
\ large }
\ \  T 
 \label{mudef}
\end{equation}
and then,  according to (\ref{proj0},\ref{Gj0}), $\mu(\lambda)$ is given by
\begin{eqnarray}
\lefteqn{\mu(\lambda)= 
 {1 \over N} \max_{j_0}[ \lambda j_0 + G(j_0)] 
=}
\label{mudef1}
\\
&{\displaystyle {1 \over N} \lim_{T \to \infty} {1 \over T}\max_{\rho(x,s)} \left[  \lambda \int_0^T
j(x,s) ds - \mathcal{I}_{[0,T]}^\nu(j,\rho)
\right]}
\nonumber
\end{eqnarray}

\bigskip

For a system of length $N$ connected to two reservoirs at densities
$\rho_a$ and $\rho_b$ at its two ends, the calculation of the large
deviation function $G(j_0)$ of the current  is therefore reduced to
finding the time-dependent profile $\rho(x,s)$ which optimizes
(\ref{Gj0}) under the constraints (\ref{conservation}) and 
(\ref{constraint}) and with the additional boundary conditions
$\rho(0,s)=\rho_a$ and $\rho(1,s)=\rho_b$. All the results of our
previous work \cite{bd}  follow then from the assumption that this optimal profile
does not vary with time (except from boundary effects near time 0 and time $T$
which do not contribute in the large $T$ limit).

For a system on a ring of $N$ sites, as we shall consider here, the
optimization problem is the same except for the boundary condition which
becomes $\rho(0,s)=\rho(1,s)$ and the fact that the total density
$\rho_0$ on the ring becomes an additional  conserved quantity
$$\int_0^1 \rho(x,s) dx = \rho_0 \, .$$

Our paper is organized as follows: in section \ref{sec: 2}, we consider  a general
lattice gas on 
a ring and show
 under what conditions the flat profile becomes unstable.
In sections \ref{sec: 2} and  \ref{sec: 3},  we write the large deviation function of the current,
when the optimal profile has a fixed shape  moving at a
constant velocity. 
In section  \ref{sec: 4}, we present exact numerical results on the weakly
asymmetric exclusion process for small system sizes which give evidence
that for some range of parameters, consistent with the results of section
\ref{sec: 3}, the optimal profile is no longer flat but becomes  space-time
dependent (for $\rho_0=1/2$ it is only space dependent). 
In section \ref{sec: 5}, we analyze the limit of a strong asymmetry and obtain  
a simple expression for the large deviation function of the current, under 
the assumption that the optimal profile becomes in this limit  a step function. 
In section  \ref{sec: 6}, we show that, even for a  strongly asymmetric case such as the totally asymmetric exclusion process,  one can exhibit time-dependent profiles 
determined by the Jensen Varadhan functional
which give, for large system size,  the exact large deviation function of
the current previously calculated by the Bethe ansatz.

\section{Conditions for the stability of a flat profile}
\label{sec: 2}

Consider a lattice gas on a ring of $N$ sites  with total density $\rho_0$.
Under the assumption that the optimal profile $\rho(x,s)$ is flat, that is
$$\rho(x,s)=\rho_0$$
the large deviation function $G(j_0)$ is given (\ref{Idef},\ref{Gj0}) by
\begin{equation}
G_{\rm flat}(j_0) = - {[j_0 - \nu \sigma(\rho_0)]^2 \over 2 \sigma(\rho_0)} 
\label{Gflat}
\end{equation}
which, by  (\ref{mudef1}), gives for $\mu(\lambda)$
\begin{equation}
\mu_{\rm flat}(\lambda) \simeq{\lambda (\lambda+ 2 \nu )\sigma(\rho_0)
\over 2 N}   \, .
\label{muflat}
\end{equation}

\bigskip

A natural question is whether one could increase $G(j_0)$  (and
$\mu(\lambda)$) by adding to this flat profile  some small ($\epsilon \ll 1$) space and time dependent perturbation of the form
$$ j(x,s) = j_0 + \epsilon [ j_1(x) \cos (\omega s) + j_2(x) \sin(\omega s) ] $$
in which case due to (\ref{conservation})
$$ \rho(x,s) = \rho_0 + {\epsilon\over \omega} [ -j_1'(x) \sin (\omega s) + 
j_2'(x) \cos(\omega s) ] $$
where $j_1(x)$ and $j_2(x)$ are periodic functions of period 1.
The resulting expression for $G(j_0)$  to second order in $\epsilon$ is
\begin{eqnarray*}
G(j_0)= - {(j_0- \nu \sigma(\rho_0))^2 \over 2 \sigma(\rho_0) } + \epsilon^2 \int_0^1 dx \left[
- {j_1^2 + j_2^2 \over 4 \sigma(\rho_0)}  \right. \nonumber \\
- {(j_1''^2 + j_2''^2) D(\rho_0)^2 \over 4 \omega^2 \sigma(\rho_0)}
+ {j_0(j_1 j_2'-j_2 j_1') \sigma'(\rho_0) \over 2 \omega
\sigma^2(\rho_0)} 
\nonumber \\ 
\left. +(j_1'^2 + j_2'^2) \left( {j_0^2 \sigma''(\rho_0) \over 8 \omega^2 \sigma(\rho_0)^2} - {\nu^2 \sigma''(\rho_0) \over 8 \omega^2} -{j_0^2 \sigma'^2(\rho_0) \over 4 \omega^2 \sigma^3(\rho_0)} \right)\right]
\end{eqnarray*}
As this expression is quadratic in the currents $j_1$ and $j_2$, the various Fourier modes are not coupled.
Choosing for $j_1(x)$ and $j_2(x)$
\begin{eqnarray}
j_1(x) = a \cos (2 \pi x) + b \sin (2 \pi x) \nonumber \\
j_2(x) = c \cos (2 \pi x) + d \sin (2 \pi x) 
\label{mode}
\end{eqnarray}
one gets
\begin{eqnarray*}
G(j_0)= - {(j_0- \nu \sigma(\rho_0))^2 \over 2 \sigma(\rho_0) } + \epsilon^2 \left[(ad-bc){j_0 \sigma'(\rho_0) \pi \over \omega \sigma^2(\rho_0) } 
\right.
\\
 - (a^2+b^2+c^2+d^2) \left( {1 \over 8 \sigma(\rho_0)} +
 {2 \pi^4 D^2(\rho_0) \over \omega^2 \sigma(\rho_0) }
\right.  
\nonumber \\
\left. \left.
 +{ \pi^2 \sigma'^2(\rho_0) j_0^2 \over 2\omega^2 \sigma^3(\rho_0) }
 +{ \nu^2 \pi^2 \sigma''(\rho_0)  \over 4\omega^2  }
 -{ \pi^2 \sigma''(\rho_0) j_0^2 \over 4\omega^2 \sigma^2(\rho_0) } \right) \right]
\end{eqnarray*}
The flat profile is stable against the perturbation (\ref{mode}), if this 
 is a negative definite quadratic form in $a,b,c,d$.
This  is achieved when for all $\omega$
\begin{eqnarray}
\lefteqn{{1 \over 8 \sigma(\rho_0)} +
 {2 \pi^4 D^2(\rho_0) \over \omega^2 \sigma(\rho_0) }
 +{ \pi^2 \sigma'^2(\rho_0) j_0^2 \over 2\omega^2 \sigma^3(\rho_0) } } \nonumber\\
& {\displaystyle +{ \nu^2 \pi^2 \sigma''(\rho_0)  \over 4\omega^2  }
 -{ \pi^2 \sigma''(\rho_0) j_0^2 \over 4\omega^2 \sigma^2(\rho_0) }   >
\left|{j_0 \sigma'(\rho_0) \pi \over 2 \omega  \sigma^2(\rho_0) }\right|}
\nonumber
\end{eqnarray}
The flat profile becomes therefore unstable if 
\begin{eqnarray}
\lefteqn{{1 \over 8 \sigma(\rho_0)} +
 {2 \pi^4 D^2(\rho_0) \over \omega^2 \sigma(\rho_0) }
 +{ \pi^2 \sigma'^2(\rho_0) j_0^2 \over 2\omega^2 \sigma^3(\rho_0) } }\nonumber\\
 &{\displaystyle +{ \nu^2 \pi^2 \sigma''(\rho_0)  \over 4\omega^2  }
 -{ \pi^2 \sigma''(\rho_0) j_0^2 \over 4\omega^2 \sigma^2(\rho_0) }   <
\left|{j_0 \sigma'(\rho_0) \pi \over 2 \omega  \sigma^2(\rho_0) }\right|}
\label{condition1}
\end{eqnarray}
i.e.
\begin{equation}
 8 \pi^2 D^2(\rho_0)   \sigma(\rho_0) +( \nu^2 \sigma^2(\rho_0) -j_0^2 ) \sigma''(\rho_0)  <0
\label{instabilite}
\end{equation}
which, given (\ref{mudef1},\ref{Gflat},\ref{muflat}), can be rewritten as
\begin{equation}
4 \pi^2 D^2(\rho_0) <  N \mu_{\rm flat} (\lambda) \sigma''(\rho_0)
\label{instabilitebis}
\end{equation}
When the flat profile becomes unstable, according to (\ref{condition1}), the current takes the form
\begin{equation}
j(x,t) = j_0 + A \cos 2 \pi\left(x - x_0 - {j_0 \sigma'(\rho_0) \over \sigma(\rho_0)} t \right) 
\label{cur-inst}
\end{equation}
where the amplitude $A$ would be determined by expanding $G(j_0)$ to higher order in $\epsilon$.

One could analyze in a similar way the stability of the flat profile
against other modes by choosing $j_1(x) = a \cos (2 \pi nx) + b \sin( 2 \pi nx) $
and $j_2(x) = c \cos (2 \pi nx) + d \sin (2 \pi nx)$
 and the threshold (\ref{instabilite}) would become $$
 8 \pi^2 D^2(\rho_0)n^2   \sigma(\rho_0) +( \nu^2 \sigma^2(\rho_0) -j_0^2 ) \sigma''(\rho_0)  <0$$
This shows that 
that the fundamental ($n=1$) is the first mode to become unstable.

\section{A simple time dependent profile}
\label{sec: 3}

The  form  (\ref{cur-inst}) suggests that beyond the instability the
optimal profile is a fixed shape moving at a constant velocity $v$
$$\rho = g(x-vt) \ .$$
Due to conservation law (\ref{conservation}) the current is  then 
$$j(x,t)=j_0 - v \rho_0 + v g(x-vt)$$
If such a profile is the optimal profile, then
the variational principle (\ref{Gj0}) reduces to
\begin{eqnarray}
\lefteqn{G(j_0) = -\min_{g(x),v} \int_0^1 {dx \over 2 \sigma(g(x)) } 
\Big[ j_0-v \rho_0 + v g(x) } \nonumber  \\
& \qquad \qquad \qquad {\displaystyle
 + D(g(x))g'(x) - \nu \sigma(g(x)) \Big]^2 }
\label{Gj0bis} 
\end{eqnarray}
This is of the form (the term linear in $g'$ gives a null contribution due to the periodic boundary conditions)
\begin{equation}
G(j_0)= - \inf_{g(x),v} \int_0^1 dx [X(g) + g'^2 Y(g)]
\label{Gj0ter}
\end{equation}
where
$$X(g)= {[j_0-v \rho_0 + v g  - \nu \sigma(g)]^2 \over 2 \sigma(g) } $$
and
$$Y(g)= {D^2(g) \over 2 \sigma(g)} . $$
The optimal $v$  in (\ref{Gj0bis}) is then given by
\begin{equation}
 v=- {\int dx {(g-\rho_0) (j_0 - \nu \sigma(g)) \over  \sigma(g) }
\over \int dx {(g-\rho_0)^2  \over  \sigma(g) }} = -j_0 {\int dx {(g-\rho_0)  \over  \sigma(g) }\over \int dx {(g-\rho_0)^2  \over  \sigma(g) }}
\label{vopt}
\end{equation}
this last simplification being due to the constraint
 $\int  g(x) dx = \rho_0$. With this constraint and for a fixed $v$, a variational calculation of the optimal $g$ in (\ref{Gj0ter})   shows that $g$ should satisfy
$$X'(g) - 2 Y(g)  g'' - g'^2 Y'(g)= C_2$$
Multiplying both sides by $g'$ allows one to integrate once so that $g$
 satisfies
\begin{equation}
X(g)-g'^2 Y(g)= C_1 + C_2 g
\label{xm}
\end{equation}
where $C_1$ and $C_2$ are constants (which is an extension of the equation (15) of \cite{bd} to the case of the ring).

For  fixed $j_0$, $\rho_0$ and $v$, if one denotes by $g_1$ and $g_2$ the two extrema of 
the profile $g$ (generically, the profile $g(x)$ is a periodic function 
of period 1 with a single 
minimum $g_1$ and a single maximum $g_2$), one can determine  the constants $C_1$ and $C_2$ by (\ref{xm}) in terms of $g_1$ and $g_2$ (as $X(g_1) = C_1 + C_2 g_1$ and $X(g_2) = C_1 + C_2 g_2$).
  The differential equation (\ref{xm})  determines  the whole  profile (up to a translation on the ring)  and the constants $g_1$ and $g_2$  are then fixed by
the fact that
\begin{eqnarray*}
\lefteqn{\displaystyle {1 \over 2}\ = \int_{x(g_1)}^{x(g_2)} dx =
\int_{g_1}^{g_2} {dg \over g'}}  \\
& &{\displaystyle  \ \ \ \ \ \ = \int_{g_1}^{g_2} \sqrt{Y(g) \over X(g) -  C_1 - C_2 g
} \; dg }
\end{eqnarray*}
and
\begin{eqnarray*}
{\rho_0 \over 2}\
 =\int_{g_1}^{g_2} g \sqrt{Y(g) \over X(g) -  C_1 - C_2 g} \; dg
\end{eqnarray*}

\section{Exact numerics for the weakly asymmetric exclusion process on a ring}
\label{sec: 4}

We wrote a program to calculate exactly $\mu(\lambda)$ for the weakly asymmetric exclusion process (WASEP) on a ring of $N$ sites with $P= N\rho$ particles.
In the simple symmetric exclusion process (SSEP), each particle  jumps to
its right at rate
${1\over 2} $
 and to its left at rate
${1\over 2} $
 and the functions $D(\rho)$ and $\sigma(\rho)$ are given 
 \cite{spohn} by
\begin{equation}
D_{\rm SSEP} = {1 \over 2}
\label{Dasep}
\end{equation}
\begin{equation}
\sigma_{\rm SSEP} = \rho(1-\rho) .
\label{sigmasep}
\end{equation}
If one introduces a weak electric field to the right, the model becomes
the WASEP and 
the rates become
${1\over 2} +{\nu \over 2 N}$ to the right
and
${1\over 2} -{\nu \over 2 N}$.

As the evolution is a Markov process, one can build, 
 as explained in \cite{dl,rdd}
from the Markov matrix,  a $\lambda$-dependent matrix, the largest eigenvalue of which is $\mu(\lambda)$ defined by (\ref{mudef}).
 According to the linear stability analysis, the flat profile becomes
unstable (\ref{instabilite}) for
$$j_0^2 <  \rho (1-\rho) [ \nu^2 \rho(1-\rho) - \pi^2 ]$$
or by (\ref{instabilitebis}) for
\begin{equation}
N\mu_{\rm flat} (\lambda) < - {\pi^2 \over 2} .
\label{pi2}
\end{equation}

\begin{figure}
\centerline{\selectlanguage{english} {\hskip1cm} \input{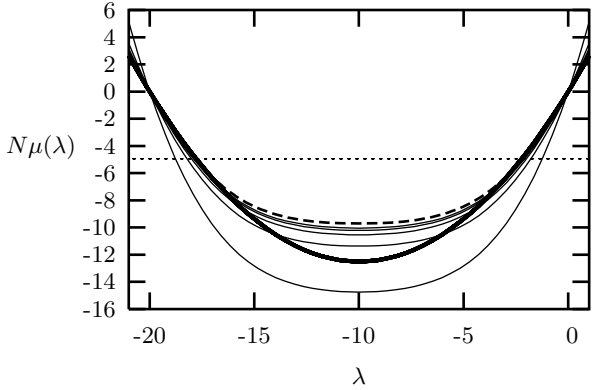}}
%\centerline{\includegraphics[width=6cm]{fig.pstex}}
\caption{\small $\mu(\lambda)$ as defined by (\ref{mudef}) 
for the weakly asymmetric
exclusion process on a ring of $N=6,10,14,18,22$ sites when the asymmetry
parameter $\nu=10$ and the density $1/2$ (thin lines). As $N$ increases, 
the data increase and seem to accumulate to the value predicted (dashed line) 
by assuming that the optimal profile is the one  discussed in section  \ref{sec: 3}. 
If the optimal profile had been flat, the curve would have been 
predicted by (\ref{muflat}) (thick line). The horizontal dotted line gives the value of $\mu$
below which the flat profile becomes unstable (\ref{pi2}). }
\end{figure}

We have calculated the exact eigenvalue $\mu(\lambda)$ for lattice sizes from $N=6$ to $22$, at  density $1/2$ for an asymmetry $\nu=10$ (in order to avoid negative rates for small system sizes, we have replaced in our programs the rates ${1\over 2} \pm{\nu \over 2 N}$ by $\exp[\pm \nu /N] /2$).
The results show a rather quick  convergence  with increasing $N$ towards
the value corresponding to a shape 
determined by (\ref{Gj0bis}). 
Clearly the flat profile gives a value too low, incompatible with the numerical data.

\section{A bridge between a weak and a strong asymmetry}
\label{sec: 5}

Assuming that the optimal profile is the one discussed in section  \ref{sec: 3}, if one 
tries to make $\nu$ large, and one writes 
\begin{equation}
j_0 = \nu i_0
\label{i0}
\end{equation}
the moving  profile which satisfies (\ref{xm}) becomes  very steep in the regions where it varies, and it takes the form of a step function with two constant values $g_1$ and $g_2$ separated by two discontinuities
\begin{eqnarray}
g(x) = g_1   \ \ \ \ {\rm for} \ \ \  0 < x < y \nonumber \\
g(x) = g_2   \ \ \ \ {\rm for} \ \ \  y < x < 1
\label{gnu}
\end{eqnarray}
so that the parameters $ g_1, g_2$ and $y$ are related to $\rho_0$ and
$i_0$
by
\begin{equation}
 \rho_0 = y g_1 + (1-y) g_2 
\label{rho0}
\end{equation}
\begin{equation}
 i_0 = y \sigma(g_1) + (1-y) \sigma(g_2)  \ .
\label{i0bis}
\end{equation}
The expression of the velocity (\ref{vopt}) then becomes
\begin{equation}
 v= \nu { \sigma(g_2) - \sigma(g_1) \over g_2 - g_1} .
\label{voptbis}
\end{equation}

Then using the fact that $g'$ vanishes when $g(x)=g_1$ or $g_2$ and 
replacing  (\ref{i0},\ref{rho0},\ref{i0bis},\ref{voptbis}) into
(\ref{xm}) implies that the constants $C_1, C_2$ (in (\ref{xm})) 
vanish at  order $\nu^2$. 
Thus asymptotically in $\nu$, one can rewrite (\ref{xm}) as
%\begin{widetext}
\begin{eqnarray}
\label{g'}
\lefteqn{g'^2 = \nu^2 { 1  \over D(g)^2 (g_1- g_2)^2} \Big[ 
g_2 (\sigma(g_1) - \sigma(g))}\\
& & \qquad \quad 
+g_1 (\sigma(g) - \sigma(g_2)) +g (\sigma(g_2) - \sigma(g_1)) \Big]^2 
\nonumber
\end{eqnarray}
and 
\begin{eqnarray}
\lefteqn{ G(j_0) = -\nu^2 \int  { dx \over  \sigma(g) (g_1-g_2)^2} 
\Big[ g_2 (\sigma(g_1) - \sigma(g)) } \nonumber\\
 & \quad 
+g_1 (\sigma(g) - \sigma(g_2)) +g (\sigma(g_2) - \sigma(g_1)) \Big]^2 .
\label{Gj0nu}
\end{eqnarray}
%\end{widetext}
If one replaces $g$ by its expression (\ref{gnu}), one gets that the
order $\nu^2$ of $G(j_0)$ vanishes. The next order in the large $\nu$
expansion is dominated by the rounding-off of the discontinuities in
(\ref{gnu}) as given by (\ref{g'}). As the profile $g(x)$ is composed
of two monotonic parts  one can then use (\ref{g'}) into (\ref{Gj0nu})
and obtain for $g_2>g_1$
\begin{eqnarray*}
\lefteqn{ G(j_0) = - 2 \nu \int_{g_1}^{g_2} dg {D(g) \over (g_2-g_1)\sigma(g)}  
\Big| \sigma(g) ( g_2 - g_1) } \nonumber \\
& & \qquad \qquad - \sigma(g_1) (g_2-g) - \sigma(g_2) (g-g_1)  \Big| 
%\label{Gj0final}
\end{eqnarray*}
It is remarkable that $y$ is not present in this expression.
In the case of the weakly asymmetric exclusion process on a ring,
expressions (\ref{Dasep},\ref{sigmasep}) of $D(\rho)$ and $\sigma(\rho)$
lead for large $\nu$ to
\begin{eqnarray}
\label{Gj0wasep}
\lefteqn{G(j_0) = -  \nu  \left[  g_2 - g_1 - g_1 g_2 \ln{g_2 \over g_1}
\right.} \\
& & \qquad \qquad  \qquad \left. 
- (1-g_1)(1-g_2 )\ln \left({1-g_1 \over 1-g_2} \right)\right]
\nonumber
\end{eqnarray}
 In this case (\ref{voptbis}) becomes $v= \nu ( 1- g_1 -g_2)$.

\bigskip

If one takes formally  $\nu=N$,  the hopping rates $1/2\pm \nu/2N$ become
$1$ and $0$, so that the model reduces to the totally asymmetric
exclusion process and
one gets from (\ref{proj0},\ref{i0},\ref{Gj0wasep})
%\begin{widetext}
\begin{eqnarray} 
\lefteqn{{\rm Pro}\left( {Q_T \over T} = i_0   \right)
\sim   \exp 
\left( -T  \left[  g_2 - g_1 - g_1 g_2 \ln{g_2 \over g_1} 
\right. \right. } \nonumber \\
& & \qquad \qquad \left. \left. 
- (1-g_1)(1-g_2 )\ln \left({1-g_1 \over 1-g_2} \right) \right] \right) \, .
\label{G1}
 \end{eqnarray}
%\end{widetext}
As we will see it  in section  \ref{sec: 6}, this is exactly the large deviation
function predicted by the Jensen-Varadhan theory \cite{JV} to maintain a profile
(\ref{gnu}) formed of a shock and an antishock in the totally asymmetric exclusion
process.
Other aspects of the relation between the large deviation functional
of the weakly asymmetric exclusion process and the Jensen-Varadhan functional
in systems with open boundary conditions will be presented in \cite{bd2}.

\section{Large deviations of the current in the totally asymmetric exclusion process}
\label{sec: 6}

In the totally asymmetric process, each particle  jumps to its
neighboring site, on its right, at rate 1, if the target site is
empty (and there is no other jump).

The large deviation function of the current of the totally asymmetric
exclusion process on a ring  of $N$ sites, with $P$ particles, 
has been calculated exactly \cite{dl,da}.
If $Q_T$ is the total number  of jumps during time $T$ over a given bond on the ring,
one knows that for large $T$, 
\begin{equation}
\langle e^{\lambda Q_T} \rangle \sim e^{\mu(\lambda) T}
\label{mudefbis}
\end{equation}
and explicit expressions of $\mu(\lambda)$ has been obtained for all $N$ and $P$ by  the Bethe ansatz \cite{dl,da}.

For large $N$, it was shown in particular (equation (53) of \cite{da} 
with the proper redefinition of the parameters) that for $\lambda <0$
\begin{equation}
\mu(\lambda) = - { (1- e^{\lambda \rho_0})(1- e^{\lambda (1-\rho_0)}) 
\over (1- e^{\lambda })} .
\label{bethe}
\end{equation}

We are going now to argue that this result can be understood, by assuming
that (\ref{mudefbis}) is dominated by configurations of the form
(\ref{gnu}) moving at a velocity $v=1-g_1-g_2$.
These density profiles are everywhere constant except for a shock (at
some position $z$ with $g(z-0)=g_1$ and $g(z+0)=g_2$ for $g_2>g_1$) 
and an antishock
(at position $z+y$ with $g(z+y-0)=g_2$ and $g(z+y+0)=g_1$).
From the Jensen-Varadhan theory \cite{JV},  the probability of maintaining such a
shape  moving at this velocity on the ring over a very long period of
time $T$ reduces to the probability of maintaining an antishock  between
the densities $g_2$ and $g_1$ 
moving at velocity $v$. The probability of the latter event,
which we denote by  $P_T (g_1,g_2)$ is given \cite{JV}  by
%\begin{widetext}
\begin{eqnarray}
\lefteqn{P_T (g_1,g_2) \sim \exp \left( -T \left[  g_2 - g_1
- g_1 g_2 \ln{g_2 \over g_1} 
\right. \right.} \nonumber \\
& & \qquad  \left. \left.  - (1-g_1)(1-g_2 )
\ln\left({1-g_1 \over 1-g_2} \right)\right] \right) .
\label{pt}
\end{eqnarray}
%\end{widetext}
The corresponding integrated current $Q_T$ is 
$$Q_T = T[y g_1 (1-g_1)+ (1-y) g_2(1-g_2)]$$
since over a long period of time $T$  a given bond spends  a fraction $y$
of the time  at density $g_1$ and $1-y$ at density $g_2$.

Therefore, if the configurations of the form (\ref{gnu}) dominate the large deviations of the current, one expects
%\begin{widetext}
\begin{eqnarray}
\lefteqn{\mu(\lambda) = \max_{y,g_1,g_2}
 \left\{  \lambda [y g_1(1-g_1) + (1-y) g_2 (1-g_2)] - 
\right.}  \nonumber \\ 
&  \left.  \left[  
g_2 - g_1 - g_1 g_2 \ln {g_2 \over g_1}
 - (1-g_1)(1-g_2 )\ln \left( {1-g_1 \over 1-g_2} 
 \right)  \right]
 \right\} \nonumber\\
& 
\label{muasep}
\end{eqnarray}
%\end{widetext}
where the maximum has to satisfy the constraint 
\begin{equation}
\rho_0 = y g_1 +(1-y) g_2
\label{cont5}
\end{equation}

A calculation of
the optimum in (\ref{muasep}), with the constraint (\ref{cont5})  leads to  
$$g_1 ={ e^{\lambda } -e^{\lambda (1-\rho_0)}  \over e^{\lambda } -1}  \ \ \ ; \ \ \ 
g_2 ={e^{\lambda \rho_0} -1 \over e^{\lambda } -1}
$$
and (\ref{muasep}) becomes (\ref{bethe}).

This shows that the result of the Bethe ansatz (\ref{bethe}) can be
physically understood in terms of an optimal profile which takes the form
of the step function (\ref{gnu}). The probability of maintaining this
profile is given by the Jensen-Varadhan expression (\ref{pt}) which in
fact is identical to (\ref{G1}) obtained, in the large $\nu$ limit, for the
WASEP from  the { hydrodynamic large deviation theory}.

That the fluctuations are due, in the strong asymmetric case, to
configurations formed by  a gas of shocks and antishocks has been already pointed
out by Fogedby \cite{f1,f2,fb}.  The calculation of this section shows  that 
the large deviation of the current, in the range $\lambda<0$,
can be understood quantitatively in terms of a single pair of shock-antishock.
Whether the Fogedby theory would allow to understand all the current
fluctuations, including the range $\lambda>0$ where the expression of
$\mu(\lambda)$ is  more complicated \cite{da} than (\ref{bethe}),
remains an interesting open question.

\section{Conclusion}
\label{sec: 7}

In the present work we have determined 
the limit of stability (\ref{instabilite},\ref{instabilitebis}) of a flat profile
{ for a diffusive lattice gas  on a ring}. 
This instability beyond which the optimal profile becomes modulated is of
the same nature as the phase transition  found for several other
non-equilibrium systems \cite{bgl,zia,ff,cde}.
As the calculation is based on a local stability analysis, one cannot of
course exclude  first order transitions, i.e. that the flat profile might become 
 globally unstable.

In section  \ref{sec: 4}, we have  obtained numerical evidence that the macroscopic fluctuation theory 
predicts correctly the large deviation function of the current for the
weakly asymmetric exclusion process. The numerical results are consistent
with the second order phase transition predicted in section  \ref{sec: 2}, and with
a modulated density profile moving at a constant velocity as suggested in
section   \ref{sec: 3}.  These results could in principle be confirmed  by solving
the Bethe ansatz equations for the WASEP, since $\mu(\lambda)$ can be calculated exactly
for the ring geometry  \cite{lk,k}.

It would be interesting to extend the results of the present work to
the case of open boundary conditions. One difficulty is that the
time-independent profile, found in \cite{bd},  is much more complicated
than the flat profile for the ring geometry, and we did not succeed so far to 
obtain the condition which would generalize
(\ref{instabilite},\ref{instabilitebis})  for this
open geometry.

Lastly, we noticed that  the large deviation function (\ref{G1})
 obtained for the weakly asymmetric diffusion process 
{ in the large drift limit}
% a case for which  the macroscopic fluctuation theory applies, 
 is identical to the one predicted for a
strong asymmetry by the Jensen-Varadhan theory (\ref{pt})
(see also \cite{bd2}).

Despite this bridge between the large deviation function of the current
of weakly and strongly asymmetric systems, and some recent results on
zero-range processes \cite{hrs},  a theory of current fluctuations for strongly
asymmetric lattice gas such as the ASEP with open boundary conditions
remains an open problem.

\section{Appendix: a  derivation of (\ref{mft},\ref{Idef})}
\label{sec: 7}

We present here an heuristic derivation of the hydrodynamic large deviations (\ref{mft}).
Let us consider a system of $N$ sites and  decompose it into $N/l$ boxes of $l$ sites each.
Let us define the density $\rho_i(t)$ in box $i$ at time $t$
and $q_i(t)$ the total number of particles transfered from box $i$ to box
$i+1$ during a  time interval $t, t+\tau$ (this time $\tau$ should be large enough for the $q_i(t)$ to be a Gaussian
characterized by its  average and  its variance as in (\ref{sigmadef},\ref{Ddef1}), but short enough compared to the characteristic time of variation of the densities $\rho_i(t)$).

If one writes that the $q_i(t)$ are Gaussian, one gets
%\begin{widetext}
\begin{eqnarray*}
{\rm Pro} \big( q_i(s),\rho_i(s) \big) 
\sim \exp \left[ - \sum_{k=1}^{T/\tau}
 \sum_{i=1}^{N/l} {1 \over {2 \sigma(\rho_i(s)) \tau \over l}} 
 \Big( q_i(s) +  \right. \\
 \left. \left.
D(\rho_i(s)){\rho_{i+1}(s)-\rho_i(s) \over l} \tau - \nu {\sigma(\rho_i(s))  \over N} \tau\right)^2  \right]
%\label{functional1}
\end{eqnarray*}
%\end{widetext}
where $k= s \tau$.
The factor $\nu/N$ comes from the weak asymmetry of the  jumps.
Clearly the conservation of the number of particles gives
\begin{equation}
\label{cons1}
\rho_i(s+\tau)= \rho_i(s)+ {q_{i-1}(s) - q_i(s) \over l}
\end{equation}
Now if one takes a continuous limit by writing
\begin{equation}
\rho_i(s) = \rho\left(i{l \over N},{s \over N^2}\right) 
\label{sca1}
\end{equation}
and one defines a rescaled current by
\begin{equation}
q_i(s) = { \tau \over N} j\left(i{l \over N},{s \over N^2}\right) 
\label{sca2}
\end{equation}
one gets
%\begin{widetext}
\begin{eqnarray*}
{\rm Pro}(j(x,s),\rho(x,s)) \sim \exp \left[ - N^{-1} \int_0^T ds 
 \int_0^1  dx 
 \right. \nonumber \\* \left.
 {\left( j(x,s) +
D(\rho(x, s)){d \rho(x,s) \over dx} - \nu \sigma(\rho(x,s))  \right)^2 
\over 2 \sigma(\rho(x,s))} \right]
%\label{functional2}
\end{eqnarray*}
%\end{widetext}
which is exactly (\ref{mft},\ref{Idef}). Furthermore
(\ref{cons1}) with the scaling (\ref{sca1},\ref{sca2}) leads
to (\ref{conservation}).

\bigskip

\noindent
{\it Acknowledgments}: 
We thank C. Bahadoran, L.  Bertini, A.  De Sole, D.  Gabrielli, G.  Jona--Lasinio, C.  Landim for very helpful discussions.

\end{document}